\documentstyle[psfig]{mn}
\newcommand{\Dla}{ Damped Lyman-$\alpha$ }

\newcommand{\beq}{\begin{equation}}
\newcommand{\eeq}{\end{equation}}
\newcommand{\noi}{\noindent}
\newcommand{\pks}{ PKS 1127-145 }

\begin{document}
\date{}
\title[ Variable 21cm absorption at $z=0.3127$]{Variable 21cm absorption at z=0.3127}
\author[Nissim Kanekar and Jayaram N Chengalur]
{Nissim Kanekar \thanks{nissim@ncra.tifr.res.in} and 
Jayaram N Chengalur \thanks{chengalu@ncra.tifr.res.in}\\
National Centre for Radio Astrophysics,\\
 Post Bag 3, Ganeshkhind, Pune 411 007, India }
\maketitle
\begin{abstract}
We report multi-epoch GMRT HI observations of the $z = 0.3127$ damped absorber towards 
the quasar \pks, which reveal variability in both the absorption profile and the 
flux of the background source, over a time-scale of a few days. 

The observed variations cannot be explained by simple inter-stellar scintillation 
(ISS) models where there are only one or two scintillating components and all of the 
ISS occurs in the Galaxy. More complicated models where there are either more 
scintillating components or some of the ISS occurs in the ISM of the $z=0.3127$ 
absorber may be acceptable. However, the variability can probably best be 
explained in models incorporating motion (on sub-VLBI scales) of a component 
of the background continuum source, with or without some ISS. 

All models for producing the variable 21cm absorption profile require small 
scale variations  in the 21cm optical depth of the absorber. The length scale for 
the opacity variations is $\sim 0.1$ pc in pure super-luminal motion models, and 
$\sim$ 10 pc in pure ISS models. Models involving sub-luminal motion, combined 
with scintillation of the moving component, require opacity variations on far 
smaller scales, $\sim $ 10 -- 100 AU.
\end{abstract}

\section{Introduction}
\label{sec:intro}

The flat spectrum radio-loud quasar \pks ($z~\sim~1.187$) is  known to vary
at many wave bands and over widely differing time scales. Long term flux monitoring 
(e.g. Wehrle et al. 1992; Bondi et al. 1994) has shown that its flux at several radio
frequencies varies on time scales ranging from days to months. The source is an
intra-day optical variable \cite{romero}. It has also been detected in the X-ray 
\cite{siebert} and the gamma-ray bands \cite{thompson}. At energies $>~100$MeV, EGRET 
observations found flux variations between the two epochs of observation 
\cite{mclaugh}.

Sub-milliarcsecond imaging of \pks at 15 GHz \cite{keller} has shown that it 
has a simple core-jet structure, with a prominent knot at the location of a sharp 
bend in the jet. This core-jet structure is also seen in VLBI observations at lower 
frequencies \cite{tingay,bondi2}. Multi-epoch VLBI observations at 1.67 GHz \cite{bondi2} 
showed a strong increase in both the flux density as well as the compactness of the core 
at  the third epoch, suggestive of the emergence of a new component. 

\Dla absorption was detected at $z \sim 0.3127$ towards \pks by Rao \& Turnshek 
(1999), using the Hubble Space Telescope. The column density of the absorbing system 
was found to be quite high, $N_{HI} = 5.1 \times 10^{21}$ per cm$^2$. Optical observations 
have so far been unable to provide an unambiguous identification of the absorber, 
as there are three galaxies close to the line-of-sight, at redshifts very similar
to that of the damped absorption \cite{lane,bb}. 21cm absorption was also detected from 
this $z=0.3127$ system by Lane et al. (1998), using the Westerbork Synthesis Radio 
Telescope (WSRT). Higher sensitivity Giant Metrewave Radio Telescope  (GMRT) observations 
\cite{ck2000} showed that the absorption profile has a number of narrow components, 
spread out over a velocity range of $\sim 100 $ km~s$^{-1}$. 

We report here on multi-epoch GMRT observations of this 21cm profile. As described
in section~\ref{sec:obs} below, the 21cm absorption profile is observed to vary on a 
timescale of a few days. This is the second extra-galactic 21cm absorber to show 
variability, the other system being the $z\sim 0.524$ damped absorber towards the 
BL~Lac object AO~0235+164 (Wolfe, Briggs \& Jauncey 1982). We discuss, in 
section~\ref{sec:discuss}, models which might produce the observed variations.

\section{Observations and Results}
\label{sec:obs}

\begin{table*}
\caption{Observing details.}
\begin{tabular}{|c|c|c|c|c|}
Observation epochs & Centre frequency & Flux density$^a$\\
(1999)& (MHz)  & (Jy)\\
 & & \\
27  Apr  & 1082.000 & $5.2\pm 0.8$ \\
6   May  & 1082.000 & $7.9\pm 1.2$ \\
10  May  & 1082.049 & $6.1\pm 0.9$ \\
10  May  & 1081.949 & $6.0\pm 0.9$ \\
17  May  & 1081.949 & $7.0\pm 0.4$ \\
22  Sep  & 1082.050 & $4.0\pm 0.6$ \\
8   Oct  & 1082.050 & $4.3\pm 0.6$ \\
10  Oct  & 1082.050 & $6.1\pm 0.9$ \\
15  Oct  & 1082.050 & $7.4\pm 1.1$ \\
\end{tabular}
\label{tab:obs}
\vskip 0.1in
${}^a$The errors quoted include systematic errors, based on earlier GMRT observations 
of standard calibrators in this observing mode.\\
\end{table*}


\pks was observed with the GMRT at 8 epochs during April -- May and September -- October 
1999. The backend used was the 30 station FX correlator, configured to  give 128 channels
over a total bandwidth of 1~MHz, i.e. a channel spacing of $\sim 7.8$ kHz (or 
$\sim 2$~km~s$^{-1}$). The number of antennas used for the observations varied 
between 7 and 12. The varying baseline coverage is, however, of no consequence since 
\pks is an extremely compact source (angular size $\la$ 15~milliarcseconds) and is 
not resolved by even the longest baselines of the GMRT. The observing details 
are given in Table~\ref{tab:obs}. 

Flux and bandpass calibration were carried out using 3C286, which was observed 
at least every 40 minutes during each observing session. The data was converted from the 
telescope format to FITS and analyzed in AIPS using standard procedures. The task UVLIN 
was used to subtract the continuum emission of the background quasar; maps were then 
produced and spectra extracted from the three-dimensional data cube. Spectra from 
different days were corrected to the heliocentric frame, scaled to the same flux level, 
and then combined to produce the final averaged spectrum. 

\begin{figure}
\centering
\psfig{file=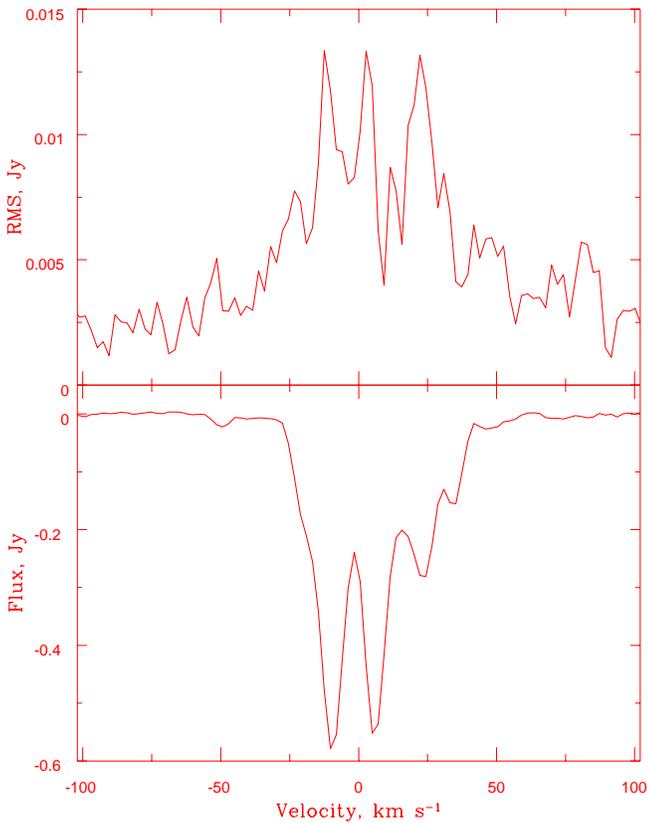,width=3.5truein,angle=0}
\caption{The lower panel shows the final averaged spectrum towards \pks
plotted versus heliocentric velocity, centred at $z = 0.3127$. The RMS noise level 
is plotted in the upper panel.}
\label{fig:rms}
\end{figure}

\begin{figure}
\centering
\psfig{file=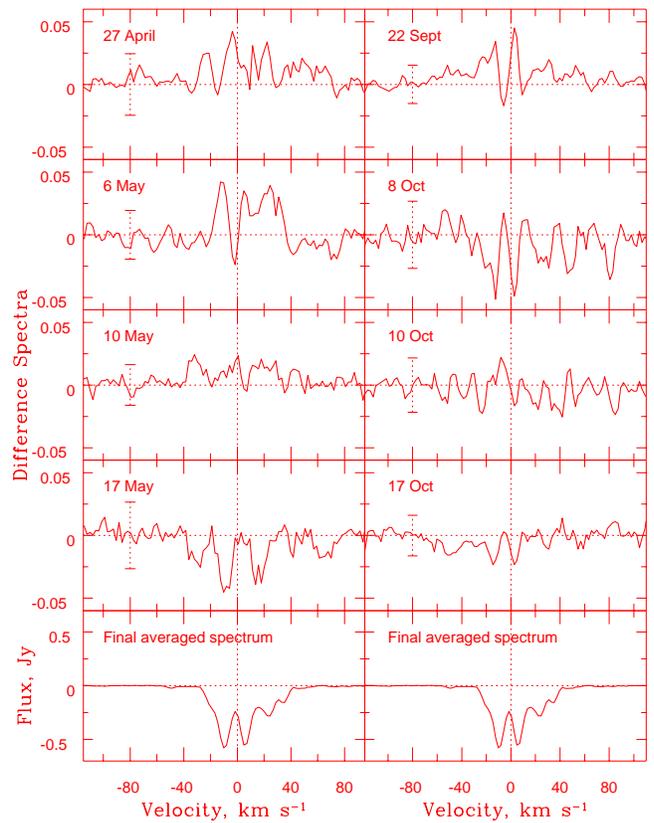,width=3.5truein,angle=0}
\caption{ Difference spectra on the 8 epochs of observation, plotted as a function 
of heliocentric velocity, centred at $z = 0.3127$. The panels are labelled by the 
dates of observation. The lowest two panels contain the final averaged spectrum, 
for comparison. Note that the scale on the y-axis is expanded for the difference 
spectra. The $3\sigma$ noise on each difference spectrum (computed in absorption-free 
regions) is indicated on the left, by the vertical barred lines.}
\label{fig:montage}
\end{figure}

The lower panel of Figure~\ref{fig:rms} shows the final spectrum, averaged over 
the 8 epochs of observation; here, flux is plotted against heliocentric velocity, 
centered at $z = 0.3127$. All spectra were scaled to a common continuum flux of 6.3~Jy,
before they were combined together, i.e. the y-axis is effectively the optical depth. 
The RMS noise level on the spectrum (note that this is computed from absorption-free 
regions in the final averaged spectrum) is $\sim 3.5$ mJy. The upper panel of the figure 
shows the RMS across the eight epochs of observation, plotted as a function of 
heliocentric velocity. It can be seen that the noise levels are far higher at line 
locations than 
in absorption-free regions, as would be expected for a time-variable absorption profile. 
Figure~\ref{fig:montage} shows a plot of difference spectra, obtained by subtracting the 
final averaged spectrum from the spectra of individual epochs (each scaled to a common 
continuum flux of 6.3~Jy; the y-axes are thus effectively the differential optical depth). 
The average spectrum is also plotted on the two lowest panels, for comparison. Note that 
the scale on the difference spectra has been expanded by a factor of 10 as compared 
to that on the average spectrum. Finally, the $3\sigma$ noise on each difference spectrum 
(computed in absorption-free regions) is indicated on the left, by the vertical barred 
lines.

An examination of figure~\ref{fig:montage} shows that the variation of different 
components is not synchronized, i.e. the observed variability is not due to some
simple problem in the scaling of the continuum flux. There is no evidence for interference 
corrupting the spectra. Since the spectra show a measurable Doppler shift over
the period of observation, an interference signal, which would stay at a fixed
frequency, would be easy to pick out. At all our observing epochs, we have
observed both polarizations; the spectra from the two polarizations (which are 
processed through  largely independent receiver chains) are identical to within
the noise. On one epoch (10th May 1999), we have two contiguous observations with the 
line profile shifted by a non-integral number of channels; the profiles
obtained from these two observations are again identical within the noise.
Finally, we have verified through simulations that the observed variations are
not the result of finite channel resolution. 

Fitting of multiple Gaussians to the average spectrum and to the spectra of 
individual epochs shows that the changes in the profile are consistent with variations
in only the depths of the various components, with no changes in their positions or 
widths. A total of nine Gaussians were fit to the final spectrum. Although the
changes in the depths of {\it some} of the components were found to be correlated 
(notably the three deepest components in Figure~\ref{fig:rms}), not all the 
components vary in a correlated manner. The observed variations in the 21cm profile of 
the $z=0.524$ damped absorber towards AO~0235+164 \cite{wolfe3} can also 
be similarly explained by changes solely in the amplitudes of individual components 
of the profile (i.e. with no changes in their centroids or widths). Note that 
although those authors quote a characteristic time scale of six months for changes 
in the $z = 0.524$ absorber, examination of Figure~(5) in Wolfe et al.~(1982) shows 
statistically significant differences between spectra taken as little as 2 days apart.

Table~\ref{tab:obs} also lists the measured flux values of \pks
for each observing session, along with error bars. The fluxes are found to vary widely,
with values ranging from $\sim 4.0$~Jy to $\sim 7.9$~Jy. The most dramatic change is 
between the 8th and the 10th of October, a flux increase of 1.8 Jy over two days. 
Our experience with the GMRT indicates that the flux calibration is reliable to 
$\sim 15$\%, in this observing mode.

\section{Discussion}
\label{sec:discuss}

As described above, the observed 21cm optical depth ($\tau_{21}$) of the 
$z = 0.3127$ absorber varies over timescales of a few days. A change in the observed 
optical depth could  arise 
from a number of reasons : (1)~changes in the physical conditions (i.e. spin 
temperature, column density) in the absorber, (2)~inter-stellar scintillation of 
one or more components of the background source and (3)~motion of a source component
across  the sky so that the line-of-sight to this component traces regions of 
different opacity. We will, in this section, discuss each of the above possibilities
in greater detail. Models for producing the variable 21cm absorption towards AO~0235+164
are discussed by Briggs (1983).

The short timescales involved make it very unlikely that the observed 
variability originates in changes in the column density along the line-of-sight, 
arising from motion of the absorbing clouds. Even if one assumes that the characteristic 
scale on which the opacity varies is 20~AU~(see below) and that the absorption 
occurs against a highly compact background source, variations over a timescale of 
2 days would require the absorbing material to have transverse velocities of $\sim 0.1c$. 
This is quite unrealistic for cold gas in a galaxy or for tidal debris. Similarly, 
there is no plausible mechanism for recurrent changes of the spin temperature of 
a cloud of this size over a timescale of days. We hence do not consider this model 
further.

Changes in the line profile could be explained if some part of the background
continuum flux arises from a region of such small angular size that it gives rise 
to inter-stellar scintillation (ISS) \cite{ger,rickett,lucyna} on passing through 
the interstellar medium of the Milky Way. The spectrum associated with this region will 
vary, while that associated with the non-scintillating flux will not; this will result 
in changes in the net optical depth. ISS will also produce variations in the continuum 
flux; further, these variations would be related to the changes in the optical depth. 
Since we measure both the background flux and $\tau_{21}$, one can test whether the 
observations are consistent with ISS models. As is discussed in more detail below, 
the lack of correlation between the variations in flux and opacity rule out the 
simplest such models.

15 GHz VLBI observations of \pks \cite{keller} have shown that it has a simple 
core-jet structure. The core is unresolved at sub-milliarcsecond resolution. The 
simplest scintillation model is that of a single compact scintillating component in 
the core, with the rest of the flux not giving rise to scintillation. In this case, 
however, two epochs with the same observed continuum flux should have the same 
observed spectrum; further, the line depths should be deeper at epochs of higher flux 
and vice versa. Table~(1) shows that the flux values agree within the error bars on 
the 22nd of September and the 8th of October; however, the spectra at these epochs are 
considerably different. Similarly, the background flux is the same on the 10th of 
May and the 10th of October while the spectra are again quite different. Given our 
error bars, the flux change required to produce the observed differences in the 21cm 
spectra would have been easily detectable, unless the line-of-sight to the scintillating 
component also coincidentally happened to have an optical depth of order unity. The 
latter situation is also, however, not viable since the epoch with the highest 
background flux ($S = 7.9 \pm 1.2$ Jy, on the 6th of May) has the weakest absorption. 
Thus, the simplest ISS  model appears to be ruled out. 

The next simplest ISS model is one in which there are two scintillating 
components and a steady component. The second scintillating component could, for 
example, arise in the bright unresolved knot in the jet of \pks, which is separated 
from the core by $\sim 2.5$ milliarcseconds. If the optical depth towards
each of the scintillating components is comparable to the average optical depth over 
the entire source (i.e. $\tau_{21} \sim 0.1$) then, as discussed below, this model is 
also ruled out by the data of 22nd September and 8th October. The difference spectrum 
on 8th October consists of two components which are $\sim$  50~mJy deeper than the 
average spectrum and three which are $\sim 30$ mJy deeper than the average. On the 22nd 
of September the first two components are weaker (by $\sim 30 - 40$~mJy) than the average,
while the remaining three agree (within the noise) with the average spectrum. If one
assumes that the absorbing clouds are distributed such that the changes in the first 
two line components arise from one of the scintillating source components (total change 
in line depth, $\Delta S \sim 90 - 100$~mJy, per component) and the changes in the 
remaining three stem from the other (total change in line depth, $\Delta S \sim 30 - 
40$~mJy, per component), the changes in line depth would correspond to a flux change 
of at least 1.3~Jy (for $\tau_{21} \la 0.1$), which is not seen (other distributions 
of the clouds would exacerbate the problem). Since these two compact
components contain far more than $10\%$ of the total flux, it is rather unlikely that
the optical depth in front of the scintillating components is $\sim 1$, while the
average optical depth is $\sim 0.1$. 

One could construct more complicated models, wherein each of the compact
VLBI components contain more than one scintillating component. The relation between the 
observed spectrum and the measured flux becomes more and more complex with an increasing 
number of scintillating components and it would hence be possible to contrive situations 
where both the variations might be explained purely as a result of ISS in
the Galaxy. However, from the Taylor-Cordes (1993) model of the distribution of 
ionized gas in the Galaxy, the scatter broadening expected towards \pks is 
0.4~milliarcseconds. It thus seems unlikely that uncorrelated ISS would result, 
even if the core or knot has multiple compact components. We note, however, that 
ISS might also occur in the galaxy(ies) giving rise to the absorption at $z=0.3127$, 
while all the above models only consider the case of it occurring in the Galaxy. 
Scintillation could prove to be a viable model if significant ISS does occur in 
the interstellar medium of the absorber.

Motion of a source component across the sky might explain the changes 
in absorption profile, if the absorbing system has opacity variations transverse 
to the line-of-sight. The line-of-sight to the moving component samples different 
paths through the absorber at different epochs, resulting in changes in the 
profile. Super-luminal motion on sub-VLBI scales has often been invoked 
to explain rapid variability of QSO fluxes \cite{rees,woltjer}. If these variations 
are intrinsic, they are difficult to explain in the standard context of incoherent 
synchrotron emission as causality arguments would imply brightness temperatures 
far in excess of the limit of $10^{12}$ K set by the inverse Compton catastrophe. 
Super-luminal motion of a source component produces a Doppler boosting of the 
observed brightness temperature by a factor $\delta^3$ over the true source 
brightness temperature, where $\delta$ is the Doppler factor. Thus, if $\delta$ 
is sufficiently large, the true brightness temperature would be lower than the 
inverse Compton limit. In these models, the timescale of variability can be 
used to place a lower limit on $\delta$. Doppler factors of order 10 are 
sufficient to account for the variability seen in the majority of compact 
sources; similar values have also been measured from VLBI observations of 
super-luminal motion. However, observed variations in the 1420 MHz flux of 
AO~0235+164 on timescales of days yield $\delta \ga 100$ \cite{krauss}. Similarly, 
long term monitoring of the flux of \pks at 4.8, 8.0 and 14.5 GHz (under the 
Michigan Monitoring program) have found short timescale variability which, if
assumed to be intrinsic,  would  require $\delta \sim 100$. Our own 
continuum measurements (albeit with larger error bars) also require $\delta$ 
values of the same order.

As mentioned earlier, the change in absorption profile in this model 
also requires the absorbing system to have opacity variations transverse to the 
line-of-sight. The extent of inhomogeneity in the absorber can be estimated 
from the fact that the physical distance $l$ travelled by the moving 
component between two observations separated by an interval $\Delta t_{obs}$ 
(in the observer's frame) is $\sim \delta c \Delta t_{obs}$, where we have 
assumed $\delta \approx \gamma$, the Lorentz factor. The corresponding projected 
distance in the absorber is $\delta c \Delta t_{obs} (D_c/D_s)$, where $D_s$ and 
$D_c$ are the angular diameter distances of the source and the absorbing clouds 
respectively. Variations in the line profile over $\Delta t_{obs} \sim 2$ days 
would require the absorbers to have opacity variations on similar scales, i.e. 
$\sim 1.6 \times 10^{-3} \delta (D_c/D_s)$ parsecs. Using the source and absorber 
redshifts in a flat, matter-dominated FRW cosmology (with $H_0 = 75$ km s$^{-1}$ 
Mpc$^{-1}$) yields $D_c \sim 775$ Mpc and $D_s \sim 1184.5$ Mpc. The absorbers must 
thus be inhomogenous on length scales of $\sim 0.1$ parsecs to explain the 
changes in the line profile in the above model. (Note, however, that the coherence 
scale of velocities in the absorber must be much larger, to account for the unchanged
positions of the velocity centroids of individual line components.) The Doppler 
factors required in this scenario are, of course, rather large but might be 
explicable in shock-in-jet models \cite{mg,qian,bk}, which are variants 
of the standard relativistic beaming paradigm.

Sub-luminal motion of a source component could also explain the 
variations in both the flux and line profile, if the moving 
component was sufficiently small so as to give rise to interstellar 
scintillation. Of course, there would be no correlation between the 
measured flux and the line profile in this situation, since the line-of-sight 
to the scintillating component would itself change. This scenario would, however, 
require the absorbing clouds to have structure on very small scales. 
For example, speeds of the order of 0.1~$c$ (at $z = 1.187$) would 
imply changes in opacity on scales of $\sim$ 20 AU, at the redshift 
of the absorber, to account for the changes in line profile over two 
days. Such fine structure in HI opacity, on scales of $10 - 100$ AU, has, 
in fact, been seen earlier in our Galaxy, in VLBI 21cm absorption studies 
\cite{faison,frail}. In fact, these observations have found the optical 
depth in the Milky Way to vary by a factor of two, over the above length scales;
this is more than sufficient to explain the changes seen in the absorption profile 
of the $z=0.3127$ absorber. We note in passing that several spin temperature  
measurements, in particular those of damped Lyman-$\alpha$ systems would 
become suspect, should 10 -- 100 AU be the typical size scale of HI ``clouds''. 
As emphasized by Deshpande (2000), however, opacity variations on scales of 
10 -- 100 AU do not neccassrily imply the existence of physically distinct 
extremely dense clouds of this size. If these variations are only random opacity 
fluctuations on small scales, with a much  weaker systematic gradient in opacity, 
the spin temperature measurements are probably reliable.

The final model we mention is one where gravitational microlensing of
a moving background component by stars or stellar clusters in the $z = 0.3127$
absorber gives rise to variations in both the flux and the line profile \cite{gk}. 
Microlensing causes the apparent transverse velocity of a moving component to
increase; this would, in turn, increase the length scale on which opacity variations
are required to exist in the absorbing source. Microlensing scenarios are strongly 
dependent on the geometry and structure of both the source and the lens. Further, 
a source which is compact enough for micro-lensing to be important would probably 
also give rise to ISS; we will hence not explore any detailed microlensing models here.
We note, however, that microlensing is a likely scenario, given the fact that 
an absorbing galaxy lies along the line-of-sight to the background quasar; simultaneous 
 multi-frequency observations could be used to test the presence of microlensing, 
due to the achromatic nature of this phenomenon.

In summary, while there are a number of models which could produce the observed
variability in both the flux and the 21cm profile, all models require the optical 
depth of the absorber to have small scale structure. The length scale for the opacity 
variations is $\sim 0.1$ pc in pure super-luminal motion models, and $\sim$ 10 pc in 
pure ISS models. Models involving sub-luminal motion, combined with scintillation 
of the moving component, require opacity variations on far smaller scales, $\sim $ 
10 -- 100 AU.
\vskip 0.1 in
\noi {\bf Acknowledgments}\\
The observations presented in this paper would not have been possible without 
the many years of dedicated effort put in by the GMRT staff to build the telescope.
We are grateful to Ger de Bruyn, K. Subramanian, Ramesh Bhat \& Frank Briggs for useful
discussions and to K. Subramanian for comments on an earlier version of this paper. 
The angular scattering towards \pks was kindly calculated by Ramesh Bhat; we thank 
him for this. This research has made use of data from the University of Michigan 
Radio Astronomy Observatory which is supported by funds from the University of Michigan.\\

\end{document}